\documentclass[aps,prd,twocolumn, showpacs, groupedaddress]{revtex4}
\begin{document}
 
%\preprint{}\sigma

\title{Absence of Right-Handed Neutrino in Weak Interactions: Explanation via Nonlinear Electroweak Model }

\author{Bill Dalton}
%\thanks{}
\affiliation{Department of Physics, Astronomy and Engineering Science,
St Cloud State University
}
\date{ \today}

\begin{abstract}
The nonlinear $SU(2)$ electroweak model is used to explain the absence of the right-handed neutrino in weak interactions. Two covariant eigenvalue constraints which affect the transformation lead to two classes of right-handed leptons, and make possible invariant mass terms without the Higgs doublet.  A covariant picture of neutrinos with mass is presented. A new invariant form for the boson potentials is described in which the boson mass terms arises via the adjoint field.  This model also indicates a different region of matter involving coupled leptons that are "blind" to the massless electromagnetic field but "see" four massive potentials that are themselves blind to the electromagnetic field. We argue that these more difficult to detect "dark" fields provide a possible contribution to the missing mass.
\end{abstract}

%\keywords{}  \mathrm{}

\maketitle
Based on work published by the author several years ago \cite{dc}, a detailed study of one particular type of nonlinear realization of Lie groups was recently presented in \cite{N}. In these realizations the presence of one field induces a nonlinear transformation on a second field. Earlier references to similar nonlinear realizations, including coset realizations, can be found in \cite{dc}, \cite{JP}, \cite{gs} and \cite{cs}. One new feature of the realizations in \cite{N} is the introduction of a covariant eigenvalue constraint on the transformations. A second feature is that the linear and nonlinear components lead to separately conserved currents for each group parameter. The three nonlinear currents reduce to a single conserved current. The latter is the electromagnetic current at one point on the adjoint sphere. These features are characteristic of the particular type of extended transformations studied in \cite{N}. 

In \cite{N} two new invariant forms involving the boson potentials were discussed for groups such as SU(n) with structure constants antisymmetric in all three indices. For nonlinear $SU(2)$ application,  it was shown that the Lagrangian for the standard gauge $SU(2) \times U(1)$ electroweak model of \cite{w} and \cite{S} is invariant under these transformations. The covariant eigenvalue conditions on the right-handed lepton component leads to two classes of right-handed leptons.These have two consequences. First, they reduce the transformation on the right-handed lepton field to a diagonal form, requiring the covariant potential for the right-handed leptons to become diagonal in one case and zero in the other. From this it follows that the coupling constants $g_V$ and $g_A$  are the same as found in the the standard gauge model. The second consequence is that the eigenvector equations require that one right-handed neutrino vanish at one place on the adjoint space $h$ unit sphere.  At this point (called the north pole in \cite{N}) the mass ratio $\frac{M_Z}{M_W} $  for the intermediate bosons take on the usual value and the mass for the $A_\mu$ field is zero. This is the only place on the $h$ sphere where this experimentally observed combination happens. At other places on the sphere the ratio $\frac{M_Z}{M_W} $ changes, the $A_\mu$ field becomes massive and this right-handed neutrino field is not zero. The second right-handed neutrino is not required to vanish at this pole point but the constraint explains why it does not participate in weak interactions. Leptons and potentials at points other than the north pole do not "see" the massless electromagnetic field. Instead, they see a heavy $A_\mu $ vector field. 

One strong experimental observation is the absence of the right-handed neutrino in weak interactions. A second is neutrino oscillations which can be explained if the neutrinos can have mass. The standard gauge $SU(2) \times U(1)$ electroweak model  accommodates, but does not explain the absence of the right-handed neutrino in weak interactions. In addition, the standard model arrives at the appropriate  $g_V$ and $g_A$ coupling constants by using different hyperfine constants for the right- and left-handed lepton components. A model that can explain both the experimentally observed absence of the right-handed neutrino in weak interactions together with the experimentally supported $g_V$ and $g_A$ coupling constants must be taken seriously. This is especially true if the same theory explains how a second right-handed neutrino can exist that does not participate in the weak interactions. This makes possible a consistent covariant picture of leptons needed to describe weak interactions and neutrino oscillations. In physics, there is an important difference between accommodating, versus explaining, experimental observation. The purpose of this paper is to describe and highlight the explanations provided by the nonlinear model of  \cite{N} for the electroweak interactions.   We give detailed expressions for both the conserved linear and nonlinear current components at the north pole point.

Following the the nonlinear realizations of $SU(2)$  in \cite{N},  transformations on the stacked spinor state $   \tilde{\mathbf{L}} = \left(\begin{array}{cc} \bar{\nu_L}& \bar{e_L}\end{array}\right)     $ have generation action given by
\begin{eqnarray}\label{Tspn}
[T_a,\mathbf{L}]     =  \frac{i}{2}\mathbf{\sigma_a} \mathbf{L} + i\xi_a\frac{1}{2}(\mathbf{-I}+\mathbf{H)}\mathbf{L}, \hspace{1.2cm} \nonumber \\\ \mathbf{H} = \left(\begin{array}{cc}h_3 \mathcal{U} & \sqrt{2}h^-\mathcal{U} \\  \sqrt{2}h^+\mathcal{U} &  -h_3\mathcal{U} \end{array}\right) ,  h^{\pm} = \frac{1}{\sqrt{2}}(h_1 \pm ih_2)
\end{eqnarray}
Here, $\mathcal{U}$ is the unit matrix in four dimensions, $Y\mathbf{L}=-1\mathbf{L}$ and the three components of $h$ transform via the adjoint representation with $[T_a,h^k] = \epsilon^{akl}h^l$and $h^kh^k = 1$. Transformation conditions for the $\xi_a$ field components are given in \cite{N}, but because these components will not enter the Lagrangian, they will not be discussed here. The three components of the $h$ field will enter the Lagrangian. In contrast to the gauge model, the group parameters for nonlinear realizations are constant with the 'local' nature of the transformation arising via the $\xi$ and $h$ fields. 

The covariant derivative acting on $\mathbf{L}  $ has the same general form as in the common electroweak gauge model. 
\begin{eqnarray}\label{P}
D_\mu \mathbf{L} = \partial_\mu \mathbf{L}   - \frac{i}{2}\mathbf{P}\mathbf{L},  \hspace{3cm} \nonumber \\\ \mathbf{P} = \left[\begin{array}{cc} (g W^3_\mu-g^\prime \beta_\mu) \mathcal{U} & g (W^1_\mu-i W^2_\mu) \mathcal{U} \\ g (W^1_\mu+i W^2_\mu) \mathcal{U} &( -g W^3_\mu-g^\prime  \beta_\mu) \mathcal{U} \end{array}\right] \hspace{.4cm}  \nonumber \\
=  \left[\begin{array}{cc} N Z_\mu \mathcal{U} & g\sqrt{2}W^- \mathcal{U} \\ g\sqrt{2}W^+ \mathcal{U} &[ -N\cos{2\theta_w} Z_\mu-2qA_\mu] \mathcal{U} \end{array}\right] 
\end{eqnarray}
Following \cite{N}, we use the standard potential relations 
\begin{equation}
\left(\begin{array}{c}W_\mu^3 \\\beta_\mu\end{array}\right)=\left(\begin{array}{cc}cos(\theta_w) & sin(\theta_w) \\-sin(\theta_w) & cos(\theta_w)\end{array}\right) \left(\begin{array}{c}Z_\mu \\A_\mu\end{array}\right),
\end{equation}
with the parameter notation $\cos(\theta_w) = \frac{g}{N}$, $\sin(\theta_w)= \frac{g^{\prime}}{N}$, $N = \sqrt{(g^{\prime})^2 + g^2}$ with the charge $q = \frac{g^{\prime} g}{N}$.
The transformations on the potentials are given by
\begin{eqnarray}\label{Twp}
[T_a,B_\mu] = \frac{1}{g^{\prime}}\partial_\mu \xi_a, \hspace{4.2cm} \nonumber \\\ [T_a, W_\mu^l]  =   \epsilon^{alk} W^k_\mu - \xi_a h^i \epsilon^{ikl} W_\mu^k +\frac{1}{g}\partial_\mu ( \xi_a h^l).
\end{eqnarray}
Notice that the action on the $W^l_\mu$  potentials has a linear and nonlinear component, where the latter involve the $\xi^a$ and $h^k$ fields. With  $\mathbf{\Gamma}^\mu= \gamma^\mu\mathbf{I}$ the Lagrangian term for the left handed lepton has the standard form 
\begin{eqnarray}\label{KL}
K_L= \frac{1}{2}[i  \tilde{\mathbf{L}}\mathbf{\Gamma}^\mu D_\mu \mathbf{L}  +(i  \tilde{\mathbf{L}}\mathbf{\Gamma}^\mu D_\mu \mathbf{L} )^\ast] 
\end{eqnarray}

With $V^k =Vh^k$ where $V$ is a group invariant, but not necessarly a space-time constant, $C_\mu^l = V^iC^{ilk}\partial_\mu V^k$ and  $C^{ijk}=\epsilon^{ijk}$ for $SU(2)$, we have from \cite{N} the following invariants
\begin{eqnarray}\label{Ka}
K_a = \big[\frac{g^2}{2}\big(V^2W^l_\mu W_l^\mu - W_\mu^l V^lW^\mu_k V^k \big) - g W_\mu^lC_l^\mu \big] 
\end{eqnarray}
\begin{eqnarray}\label{Kb}
F_\mu= g W_\mu^l V^l- \beta_\mu V g^{\prime}, \hspace{.5cm}  K_b = \frac{1}{2} F_\mu F^\mu 
\end{eqnarray}
Notice that the individual vectors $F_\mu$ are invariant under $SU(2)$. The quadratic forms $K_a$ and $K_b$ are invariant under both $SU(2)$ and the Lorentz group. 

In the limit that $V_1 \to 0,V_2 \to 0$ the expression involving the $A^\mu A_\mu$  factor vanishes and the invariant $K_a+K_b$ reduces to
\begin{eqnarray}\label{Kbb}
 K_a+K_b\to \frac{V^2}{2}(2 g^2 W_\mu^- W_+^\mu +N^2 Z^\mu Z_\mu ).
\end{eqnarray}
The space-time dependence of $ V = V_3 $ at this north pole is not specified, but in the constant limit $V \to \nu_0 /2 $, the invariant (\ref{Kbb}) at the north pole, has the same form as the expression involving the $W_\mu^{\pm} $ boson and $Z_\mu$ mass terms obtained in the standard model \cite{N}. In this alternate model the source of the mass for the bosons is shifted from the doublet to the adjoint field.  A potential $U(V)$ such as the hat potential may be added to "explain" the constant limit for $V_3$ in lowest order. The physics question is "Are the masses of the intermediate bosons constant throughout the universe?" There is insufficient data to answer this question. However, with the possible association made in \cite{N} of the non north pole region with dark matter, the space-time dependence of $V$ would be directly linked to the distribution of dark matter.

We emphasize that at this north pole point, the $A_\mu$ becomes massless, but only at this point. At other points on the sphere the $A_\mu$ field becomes massive. Fields in this non north pole domain are disjoint from fields at the north pole. They "see" a massive $A_\mu$ field. Starting from the north pole, a rotation to the non pole domain would mean the presence of $V^{\pm}$ fields. The very presence of these give mass to the $A_\mu$ field. The non pole domain represents a vast source of massive leptons and bosons. It was pointed out in \cite{N},  that the fields in this domain could provide a significant contribution to the missing mass. With (\ref{Twp}), the standard Yang-Mills \cite{YM} field expressions are covariant. These are used in \cite{N} for both the standard and alternate Lagrangians to incorporate the kinetic terms for the boson fields.The invariant (\ref{Ka}) contains a first order kinetic term for the $V^k$ components. An invariant quadratic kinetic terms for the $V^k$ components is
\begin{eqnarray}
K_V = \frac{1}{2}(\partial^\mu V_k )\partial_\mu V^k.
\end{eqnarray}

To complete the construction of the invariant lepton Langrangian, we consider an alternate formulation for which $Y\mathbf{R} = -1\mathbf{R}$ where $ \mathbf{R}$ is the right-handed lepton field with $  \tilde{\mathbf{R}} = \left(\begin{array}{cc} \bar{\nu_R}& \bar{e_R}\end{array}\right)   $.
Initially the transformations on $ \mathbf{R}$  are expressed exactly like those for $\mathbf{L} $. 
\begin{eqnarray}\label{TRs}
[T_a,\mathbf{R}]     =  \frac{i}{2}\mathbf{\sigma_a} \mathbf{R} + i\xi_a\frac{1}{2}\big(-\mathbf{I}+ \mathbf{H}\big)\mathbf{R} 
\end{eqnarray}
Following \cite{N}, we impose the matrix eigenvalue constraint $\mathbf{H}\mathbf{R}=\mathbf{h} \cdot \mathbf{\sigma} \mathbf{R} = \lambda \mathbf{R}$.
The eigenvalues are $\lambda_{\pm} = \pm h $ where $  h = \sqrt{h^k h^k} = 1$.
This matrix eigenvalue equation is covariant under the group for either of the eigenvalues. For the eigenvalue $\lambda = -1$ the transformation generators reduce to the form
\begin{eqnarray}\label{TRs2}
[T_a,\mathbf{R}^-]     =  \frac{i}{2}\mathbf{\sigma_a} \mathbf{R}^- - i\xi_a\mathbf{R}^-, 
\end{eqnarray}
with a diagonal local nonlinear part. Notice that we got a ($-1$) from the $Y\mathbf{R^-} = -1\mathbf{R^-}$ condition, and a second ($-1$) from the eigenvalue $\lambda=-1$, giving a net factor of (-2).  In the gauge picture, ( with the notation of \cite{N} ) this factor is obtained by imposing the condition  $Ye_R = -2 e_R$.  The covariant derivative becomes
\begin{eqnarray}\label{D:3}
D_\mu^-\mathbf{R}^-  = \partial_\mu \mathbf{R}^- +  iB_\mu g^{\prime} \mathbf{R}^- \nonumber \\\
g^{\prime}B_\mu =  -N\sin{\theta_w}^2 Z_\mu +qA_\mu 
\end{eqnarray} 
Except that $\mathbf{R}^-$ includes the right-handed neutrino component, this expression has the same form as the covariant derivative for the "singlet"  component of the standard electroweak model. 

To confront the presence of the right-handed neutrino component, we look at the eigenvector equations for $\lambda = -1$.
\begin{eqnarray}\label{C1}
(1+h_3)\nu_R^- + \sqrt{2}h^-e_R^-= 0, \nonumber \\\ \sqrt{2}h^+\nu_R^- + (1-h_3)e_R^- = 0
\end{eqnarray}
These two equations require that the right-handed neutrino $\nu_R^-$ vanish at the north pole $h_3=1$. This is the point where the $A_\mu$ field becomes massless and the intermediate boson mass ratio $\frac{M_Z}{M_W} $ takes on the observed value. This is one reasonable explanation of the observed absence of the right-handed neutrino in weak interactions. This is important, especially when combined with the fact that it also reduces the covariant derivative term for the right-handed lepton to the diagonal form needed to give the  $g_V$ and $g_A$ relations that are consistent with observation. In this picture, most of our experiments involving charged particles takes place at this north pole, since this is the only place on the sphere where we have the massless electromagnetic field. The constraint (\ref{C1}) does not mean that  $\nu_R^-$ vanish at other points on the $h$ sphere. 

Motivated by observations of neutrino oscillations we consider the second eigenvalue constraint ($\lambda=+1$). For this case we have 
\begin{eqnarray}\label{TR2}
[T_a,\mathbf{R}^+]     =  \frac{i}{2}\mathbf{\sigma_a} \mathbf{R}^+, \hspace{.3cm} D_\mu\mathbf{R}^+  = \partial_\mu \mathbf{R}^+ 
\end{eqnarray}
For $\lambda = +1$ the eigenvector  constraint is 
\begin{eqnarray}\label{C2}
(-1+h_3)\nu_R^+ + \sqrt{2}h^-e_R^+ = 0, \nonumber \\\ \sqrt{2}h^+\nu_R^+ - (1+h_3)e_R^+ = 0.
\end{eqnarray}
At the north pole, these equations require that $e_R^+ \to 0$, but give no restriction on $\nu_R^+$. Notice that in the $\lambda=+1$ case the right-handed components have no covariant potential term. This means that $\mathbf{R}^+$ plays
no role in weak decay at the north pole. This is consistent with observation.

For each eigenvalue case we have the following invariant form
\begin{eqnarray}\label{KR}
KR= \frac{1}{2}[i  \tilde{\mathbf{R}}\mathbf{\Gamma}^\mu D_\mu \mathbf{R}  +(i  \tilde{\mathbf{R}}\mathbf{\Gamma}^\mu D_\mu \mathbf{R} )^\ast] ,
\end{eqnarray}
where in each case the appropriate diagonal covariant derivative discussed above is used.
We also have for each case the invariant lepton mass forms.
\begin{eqnarray}\label{Km}
Km = -m [ \tilde{\mathbf{L}}\mathbf{R} + \tilde{\mathbf{R}}\mathbf{L} ] \nonumber \\  = -m[\bar{\nu_R}\nu_L +  \bar{\nu_L} \nu_R]  -m[\bar{e_R}e_L +  \bar{e_L} e_R ].
\end{eqnarray} 
Here $m$ is an invariant used to represent the mass of the lepton field. The reader should recall that in the standard model the lepton mass term involved a product of a constant times a Higgs doublet component. Here, we could express the mass as a product like $m = GV$ where $G$ is a constant. This product form is not needed for invariance of the Lagrangian, but may be introduced for other reasons.
At the north pole  $\nu_R^- \to 0$ and  $e_R^+ \to 0$. At this point the combined right-handed kinetic Lagrangian term reduces to 
\begin{eqnarray}
KR^- + KR^+ =\to \frac{i}{2}[\bar{e}_R^- \gamma^\mu \partial_\mu e_R^- -(\partial_\mu \bar{e}_R^-)\gamma^\mu e_R^-] \nonumber \\\
-B_\mu g^{\prime}\bar{e}_R^-\gamma^\mu e_R^- + \frac{i}{2}[\bar{\nu}_R^+ \gamma^\mu \partial_\mu \nu_R^+ -(\partial_\mu \bar{\nu}_R^+)\gamma^\mu \nu_R^+] 
\end{eqnarray}
The corresponding combined mass term becomes
\begin{eqnarray}
Km^- + Km^+ \to -m^-[\bar{e}_R^- e_L +\bar{e}_L e_R^-] \nonumber \\\ 
-m^+[\bar{\nu}_R^+ \nu_L +\bar{\nu}_L \nu_R^+]  
\end{eqnarray}
The field equations become
\begin{eqnarray}
-i\gamma^\mu \partial_\mu  e_R^- + g^{\prime}B_\mu \gamma^\mu e_R^- + m^-  e_L=0    \\\   -i\gamma^\mu \partial_\mu  e_L -\frac{1}{2}\big[P^{21}_\mu \gamma^\mu  \nu_L + P^{22}_\mu \gamma^\mu e_L\big] +m^- e_R^-=0    \\\
-i\gamma^\mu \partial_\mu  \nu_R^+ + m^+\nu_L=0  \\\ -i\gamma^\mu \partial_\mu  \nu_L -\frac{1}{2}\big[P^{11}_\mu \gamma^\mu  \nu_L + P^{12}_\mu \gamma^\mu e_L\big] + m^+ \nu_R^+ =0
\end{eqnarray}
In the limit that the potentials vanish, $m^+$ becomes the mass of the free neutrino and $m^-$ becomes the mass of the free electron. 

The three linear and nonlinear components of these transformations have separately conserved currents. In \cite{N} the three nonlinear currents reduced to a common conserved current $j_\mu$. At the north pole for the standard Lagrangian with $Ye_R = -2e_R$ and the transformations given in \cite{N}, this current is 
\begin{eqnarray}\label{NP}
j^\mu = \bar{e_L}\gamma^\mu e_L + \bar{e_R}\gamma^\mu e_R -i\big[W_-^{\mu\rho}W^+_\rho- W_+^{\mu\rho}W_\rho^-\big]  \nonumber \\\ +i\big[(\partial^\mu \Phi_1^\ast)\Phi_1 - (\partial^\mu \Phi_1)\Phi_1^\ast \big] \hspace{2cm}  \nonumber 
\\\ +\big[N\cos{2\theta_w} Z_\mu + 2qA_\mu\big]\Phi_1^\ast\Phi_1\hspace{2cm} \nonumber \\\ -\frac{g}{\sqrt{2}}\big[W_+^\mu\Phi_2^\ast \Phi_1 + W_-^\mu\Phi_2\Phi_1^\ast \big]. \hspace{2cm}
\end{eqnarray}  
This current at this pole is proportional to the electromagnetic current density. The usual gauge theory practice is to set $\Phi_1=0$ and $\Phi_2= \frac{\nu_o}{\sqrt{2}}$, a constant, in lowest order to generate the intermediate boson masses with the Higgs doublet. Recall that here we are using $SU(2)$ realized nonlinearly versus a gauge $SU(2)\times U(1)$. In this picture the nonlinear components generate the electromagnetic current conservation at the north pole.

For the alternate Lagrangian we drop the terms involving the Higgs doublet $\Phi$. The linear transformations of the $V$ space components give no contribution to the nonlinear current.  With  $YR^- = -1R^-$ we have
\begin{eqnarray}\label{NPa}
j^\mu = \bar{e_L}\gamma^\mu e_L + \bar{e}_R^-\gamma^\mu e_R^- -i\big[W_-^{\mu\rho}W^+_\rho- W_+^{\mu\rho}W_\rho^-\big] 
\end{eqnarray}  
For the alternate Lagrangian the linear currents at the north pole become
\begin{eqnarray}\label{JL}
J^\mu_a = -\frac{1}{2}\big[(\bar{e}_L\gamma^\mu e_L + \bar{e}_R^-\gamma^\mu e_R^-)\sigma_a^{22} \nonumber \\\
+(\bar{\nu_L}\gamma^\mu \nu_L + \bar{\nu}_R^+\gamma^\mu \nu_R^+)\sigma_a^{11} \nonumber \\\ +\bar{\nu}_L\gamma^\mu e_L\sigma_a^{12} + \bar{e}_L\gamma^\mu \nu_L\sigma_a^{21}\big]  \nonumber \\\  -\big[W_-^{\mu\rho}(\epsilon^{a1k} + i\epsilon^{a2k})W_\rho^k  +W_+^{\mu\rho}(\epsilon^{a1k} - i\epsilon^{a2k})W_\rho^k \big] \nonumber \\\
+   \cos(\theta_w) \epsilon^{a3k}W_\rho^k  \big[-Z^{\mu\rho} \nonumber \\+ iN\cos^2(\theta_w)[W^\mu_+W^\rho_--W^\rho_+W^\mu_-] \big]\nonumber \\\ + \sin(\theta_w) \epsilon^{a3k}W_\rho^k  \big[-F^{\mu\rho} \nonumber \\ +iq[W^\mu_+W^\rho_--W^\rho_+W^\mu_-]\big]
\end{eqnarray} 
Recall that at this north pole point $e_R^+ \to 0$ and $\nu_R^- \to 0$, so these fields do not appear in the conserved currents at the north pole. Even when the potential contributions to the conserved linear currents vanish, the right-handed neutrino $\nu_R^+$ still contributes to the conserved currents. The conserved transition currents $J^\mu_1$ and  $J^\mu_2 $ involve cross terms between the left-handed electron and neutrino fields.
There is a fundamental difference in the conserved currents for the standard and alternate Lagrangian given in \cite{N}. In the standard Lagrangian, $e_R$ transforms as a singlet and there is no $\nu_R$ component. The conserved linear current is the same as (\ref{JL}) if we drop the right handed components, and add the following Higgs doublet contribution.
\begin{eqnarray}
J^\mu(\Phi) = \frac{i}{2}\big[(D_\mu\Phi)^\dagger\Phi - \big((D_\mu\Phi)^\dagger\Phi\big)^\dag\big] 
\end{eqnarray}
The standard covariant derivative for $\Phi$ is given in \cite{N}.
The  currents in the above expressions are for the north pole point only. At non pole points on the sphere none of the lepton components vanish. Before the field equations and conserved currents can be obtained, the constraint equations must be incorporated into the Lagrangian. 

Since this nonlinear theory is consistent with observations at the north pole point, it is reasonable to think that the other regions on the adjoint sphere likewise exist in nature. At all regions on the sphere other than the two poles, both lepton components in each eigenvalue case have mass. All four boson potential fields are very massive. The mass of the $A_\mu$ field decreases to zero as one approaches the north pole, but is very heavy near the south pole.  This large region between the poles may be difficult to access in the laboratory because in this region interactions are with four massive potentials. Because of the large masses, interactions would perhaps be fast. The mass of the leptons and four heavy boson fields in the region between poles could provide a significant contribution to the missing mass. It is perhaps incorrect to call the leptons, $V$ space components and vector bosons at points other than the north pole, "dark" matter. They simply do not interact with the massless electromagnetic field that exists only at the north pole in this picture. Observation will depend on appropriate detectors for this region. For instance, the $e$ fields in this region would not be seen as curved tracks in a magnetic field, nor affected by electromagnetic accelerators, no matter how high the energy. At present, detection is indirectly via gravity.
\begin{acknowledgments}
The author would like to think Professor Kevin Haglin for reading this manuscript and for many useful discussions on the variety of topics raised in this study. 
\end{acknowledgments}


\begin{thebibliography}{44}
\bibitem{dc} B. J. Dalton, International Journal of Theoretical Physics {\bf 21}, 765 (1982). 
\bibitem{N} Bill Dalton, Submitted for publication. The preprint can be found at arXiv:1005.2384v1 [hep-th].
\bibitem{JP} B. J. Dalton, J. Math. Phys. {\bf 20}, 1520 (1979).
\bibitem{gs} B. J. Dalton, International Journal of Theoretical Physics {\bf 23}, 765 (1984).
\bibitem{cs} B. J. Dalton, J. Math. Phys. {\bf 19}, 1335 (1978).
\bibitem{w} S. Weinberg, Phys. Rev Lett. {\bf 19}, 1264 (1967).
\bibitem{S} A. Salam, Elementary Particle Theory, ed. N. Svaratholm, Stockholm: Almquist and Forlag, (1968).
\bibitem{YM} C. N. Yang and R. L. Mills, Phys. Rev. {\bf 96}, 191 (1954)
\end{thebibliography}
\end{document}